\documentclass[twocolumn, twocolappendix]{aastex631}
\let\tablenum\relax
\usepackage{amsfonts,amsmath,graphicx,natbib,url,hyperref,nicefrac,siunitx}
\usepackage{amssymb,verbatim,paralist,soul,comment,xcolor}

\definecolor{HexPink}{HTML}{CB1C8D}
\hypersetup{colorlinks=true, linkcolor=HexPink, urlcolor=HexPink, citecolor=blue}

\usepackage{color}

\begin{document}

\title{Double Detonations on a Moving Mesh with \texttt{Cattails}: Characterizing Early Reddening in SNe Ia}

\author{Dillon L. Hasenour}
\affiliation{Department of Physics and Astronomy, Purdue University, 525 Northwestern Avenue, West Lafayette, IN 47907, USA}

\author{Paul C. Duffell}
\affiliation{Department of Physics and Astronomy, Purdue University, 525 Northwestern Avenue, West Lafayette, IN 47907, USA}

\author{Abigail Polin}
\affiliation{Department of Physics and Astronomy, Purdue University, 525 Northwestern Avenue, West Lafayette, IN 47907, USA}

\begin{abstract}

Type Ia supernovae are useful as cosmological distance indicators and are significant sources of chemical enrichment in galaxies, but much about their origins remains uncertain. The double detonation explosion mechanism is one promising model for these transients. We present a suite of thermonuclear hydrodynamic double detonation models of sub-Chandrasekhar mass white dwarfs spanning thin to thick helium shells. Thin shell models produce light curves and spectra consistent with normal Type Ia supernovae, and thicker shell models have early time luminosity bumps from radioactive ashes in the shell and early reddening in color sometimes observed in peculiar Type I supernovae. The reddening is due to line blanketing from heavy elements produced in the shell detonation encompassing the central emission. We perform the computations using \texttt{Cattails}, a moving mesh nuclear hydrodynamics code introduced in this work. After carrying out subsequent radiative transfer calculations using the \texttt{Sedona} code, we relate the time of early reddening to the mass of the initial helium shell. We also consider how the relation changes as a function of total mass as well as how the peak brightness depends on the total mass. Finally, we provide a series of fitting functions that can be used to estimate the mass of the shell consistent with an observation of reddening in the early color evolution of a type Ia supernova. 

\end{abstract}


\section{Introduction} \label{sec:intro}

The specific progenitor scenario and explosion mechanism of Type Ia supernovae (SNe Ia) remains uncertain. One promising model, the double detonation, concerns a sub-Chandrasekhar mass carbon/oxygen (C/O) white dwarf (WD) with a shell of helium material accreted from a companion \citep{Nomoto_1982, Livne_1990, Woosley_1994}. If the helium shell begins thermonuclear runaway, it can launch a burning front wrapping around the surface and sending a shockwave into the core. This converging shockwave can then ignite the C/O in a second denotation which burns and unbinds the material and gives rise to a supernova.

Many different observations have supported the double detonation explosion mechanism. For instance, SN 2018byg \citep{De_2019}, SN 2016hnk \citep{Jacobson-Gal_2020}, SN 2022joj \citep{Liu_2023}, SN 2018aoz \citep{Ni_2023}, and SN 2020jgb \citep{Liu_2023b} are thermonuclear supernovae for which a double detonation origin has been proposed to explain early excesses or reddening. As these events span a range of luminosities and evolution rates, this suggests the mechanism operates at a range of masses; however, it can be difficult to distinguish from other scenarios in the normal Ia population. \cite{Polin_2019} shows evidence for thin shell models to exhibit a mass-luminosity-velocity relationship (where brighter supernovae/initially more massive white dwarfs have faster Si II absorption features at peak) which explains a subpopulation of SNe Ia covering normal SNe Ia and provides one method to identify double detonations in the population. Further, Gaia observations of hyper-velocity white dwarfs are associated with a double degenerate, double detonation scenario with a surviving companion \citep{Shen_2018}. 

Additionally during the infant phase of SNe Ia (the earliest light of the event), observed colors are initially diverse and populations emerge including early-red and early-blue behavior \citep{Stritzinger_2018, Ni_2025}. This hints at the possibility for distinct progenitor channels or for the wide behavior to be the extremes of one channel. Double detonations are a channel that can exhibit both early-red and early-blue behavior \citep{Polin_2019}, and further a few observed SNe Ia have early time reddening ``bumps" before the peak of emission comparable to reddening in intermediate shell double detonation models \citep{Bulla_2020}. 

The double detonation scenario can cause early reddening of emission as a result of line blanketing of the bluer wavelengths from interactions with the helium shell ashes. Thin shell models (i.e. less massive shell models) produce little to no intermediate mass elements (IMEs) or above in the shell as the burning occurs at low densities, and so they result in little to no reddening. These models agree well with normal SNe Ia \citep{Shen_2018a, Townsley_2019}. On the other hand, the thicker shell models (i.e more massive shell models) burn at higher densities to heavier elements including IMEs and radioactive isotopes. Heating from decays in the shell can result in the earliest emission being very blue, but this is immediately followed by a period of reddening as the heavy elements blanket the bluer emission from the core \citep{Polin_2019}. One aim of this work is to perform the first systematic parameter survey characterizing this early time reddening to shell mass using a new suite of double detonations.

Many recent works build on our understanding of the double detonation mechanism through two and three dimensional (2D and 3D) numerical calculations using a combination of nuclear hydrodynamics codes and radiative transfer codes \citep{Gronow_2020, Boos_2021, Pakmor_2022, Boos_2024, Zingale_2024}. However, due to the computational expense of the higher dimensions, one-dimensional (1D) models remain a useful tool in conducting parameter surveys over a wide range of initial conditions and/or finely spaced parameters. This work sets out with two primary aims: introduce a new spherically symmetric, moving mesh, nuclear hydrodynamics code called \texttt{Cattails} and build out a parameter suite of 1D double detonations characterizing the relationship between helium shell mass and early time reddening.

As a code introduction, we describe in depth the numerical methodology of \texttt{Cattails} in $\S$\ref{sec:method}. We detail the process for constructing and igniting the initial white dwarfs in $\S$\ref{sec:initial_cond}. From explosion to homologous expansion, we present the hydrodynamical models in $\S$\ref{sec:models}. Additionally, we summarize the synthetic observables from radiative transfer calculations in this section. In $\S$\ref{sec:results}, we analyze the relationships between total mass, shell mass, and reddening time. Finally in $\S$\ref{sec:discussion}, we summarize and discuss.

\section{Numerical Methods} \label{sec:method}

\texttt{Cattails} is an extension of the one-dimensional, moving mesh hydrodynamics code \texttt{RT1D} including the reaction network methodology described in \cite{Hasenour_2025}. \texttt{RT1D} is used as a base for the moving mesh, spherical geometry and careful treatment of self-gravity \citep{Duffell_2016}. \texttt{Cattails} will be publicly available with ongoing development on \texttt{Github} at \url{https://github.com/DHasenour/Cattails}, and a frozen version of the code used in this work is available at the \texttt{Zenodo} page \citep{Cattails_suite}. 

\subsection{Strang Splitting}
The overall numerical scheme can be organized into a set of conservation laws in mass, momentum, and energy that evolve a given state forward in time according to both hydrodynamical terms and reaction network terms. To combat the different timescales of the two sources, we are using a Strang split approach \citep{Strang_1968}. For a given timestep, we implicitly evolve the conserved quantities for half the total time according to reactions, then explicitly evolve for the full time according to hydrodynamics, and finally implicitly evolve the remaining half-step according to reactions. This scheme effectively centers the reactions in time over one timestep to achieve second order coupling. Symbolically,
\begin{equation}
    \label{eq:Strang_splitting}
    \begin{gathered}
        \left[\frac{\partial \mathbf{U}^n}{\partial t} = \mathcal{R}(\mathbf{U}^n)\right]_{\Delta t/2} \longrightarrow \mathbf{U}^* \\
        \left[\frac{\partial \mathbf{U}^*}{\partial t} = \mathcal{A}(\mathbf{U}^*)\right]_{\Delta t} \longrightarrow \mathbf{U}^{**} \\
        \left[\frac{\partial \mathbf{U}^{**}}{\partial t} = \mathcal{R}(\mathbf{U}^{**})\right]_{\Delta t/2} \longrightarrow \mathbf{U}^{n+1},
    \end{gathered}
\end{equation}
where $\mathbf{U}^n$ is the state at timestep $n$, $\mathcal{A}(\mathbf{U})$ are terms due to hydrodynamics, $\mathcal{R}(\mathbf{U})$ are terms due to reactions, $\mathbf{U}^*$ and $\mathbf{U}^{**}$ are intermediate states, and $\Delta t$ is the length of the global timestep. 

\subsection{Hydrodynamics}
Each hydrodynamical step, \texttt{Cattails} solves the one-dimensional, spherical Euler's equations in conservation law form:
\begin{equation}
\label{eq:convo_laws}
    \begin{gathered}
        \partial_t(\rho) + r^{-2}\partial_r (r^2 \rho v ) = 0 \\
        \partial_t(\rho v) + r^{-2}\partial_r (r^2 (\rho v^2 + P)) = 2P/r + S^{(g)}_P\\
        \partial_t\left( \frac{1}{2}\rho v^2 + \epsilon \right) + r^{-2}\partial_r \left( r^2\left[ \frac{1}{2}\rho v^2 + \epsilon + P \right]v \right) = S^{(g)}_{\epsilon} \\
        \partial_t(\rho X_i) + r^{-2}\partial_r( r^2 \rho X_i v ) = 0,
    \end{gathered}
\end{equation}
where $\rho$, $v$, $P$, $\epsilon$, and $X_i$ are the primitive variables and stand for density, velocity, pressure, internal energy density, and mass fraction of isotope $i$ respectively. During this step, the masses of individual isotopes are conserved and the mass fractions are treated like passive scalars. The energy release from nuclear reactions is captured by including the total binding energy of our composition into the total internal energy:
\begin{equation}
\label{eq:e_int}
    \epsilon = \epsilon^{(\rm eos)}_{\rm th} (\rho,P,\mathbf{X}) + \sum_i \rho E_{\mathrm{b},i} X_i,
\end{equation}
where $\epsilon^{(\rm eos)}_{\rm th}$ is the thermal energy density from the equation of state, and $E_{\mathrm{b},i}$ is the specific internal energy of isotope $i$. As the composition changes during reaction steps, energy passes between binding energy and thermal energy. We close the equations with the stellar equation of state \texttt{helmeos} \citep{Timmes_2000}. Under these assumptions, the fluid is composed of ideal gas ions, degenerate/relativistic electrons, and radiation, all in local thermodynamic equilibrium (LTE).

\begin{figure}
\centering
\includegraphics[width=0.4625\textwidth]{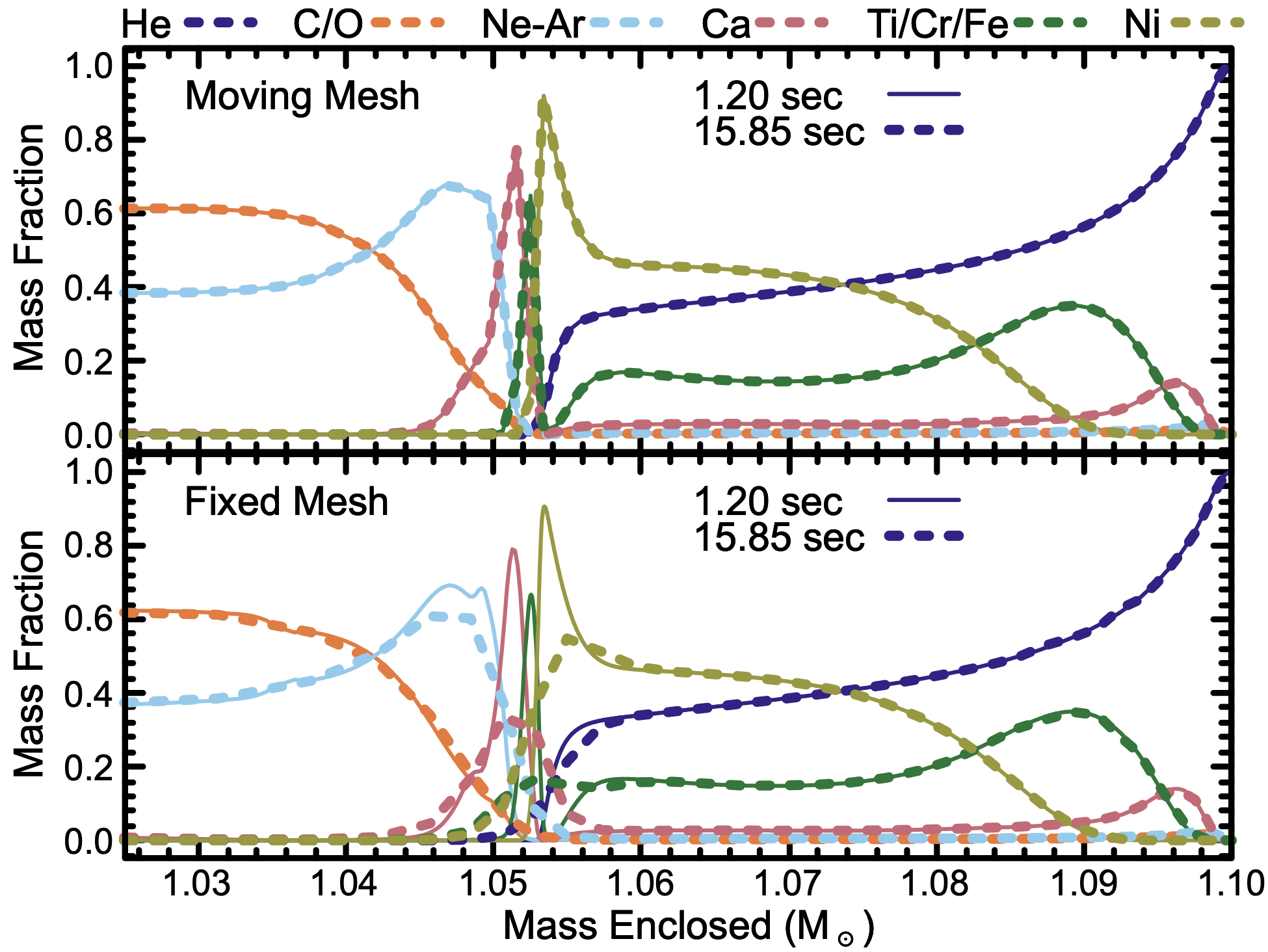}
\caption{The composition profiles for a double detonation with total mass \SI{1.1}{M_{\odot}} and shell mass \SI{0.05}{M_{\odot}} using both a moving mesh and a fixed mesh. The nuclear reactions are effectively frozen out by \SI{1.20}{sec}, but the fixed mesh case shows the solution spreading over time due to numerical diffusion. The moving mesh run greatly reduces this artificial diffusion as less material transfers between zones during this phase of expansion.}
\label{fig:FIXvsMOV.png}
\end{figure}

In practice, \texttt{Cattails} advances the computational domain by arranging it into mesh of spherical shells and using the integral form of Equation \ref{eq:convo_laws}. Once during each timestep, the moving mesh method prescribes velocities for the interfaces between neighboring zones according to the local fluid velocity. This in turn requires we subtract off the conserved material overtaken by the interface from the flux, because that material remains in the original zone. Therefore, the hydrodynamical update for each conserved quantity now takes the following form:
\begin{equation}
\label{eq:timestep}
    \begin{gathered}
    M^{n+1}_i = M^{n}_i - \Delta t\big[\Delta A (F-wU)\big]_{i+1/2} \\
     + \Delta t\big[\Delta A (F-wU)\big]_{i-1/2} ,
    \end{gathered}
\end{equation}
where $M^{n}_i$ is the conserved quantity in zone $i$ at timestep $n$, $\Delta t$ is the length of the timestep, and the left and right interface quantities of flux $F$, velocity $w$, and conserved quantity $U$ are denoted by $i\pm 1/2$. In 1D spherical geometry, $\Delta A$ is simply the spherical surface area of the interface, and $w$ is the interface velocity, chosen by averaging the fluid velocity on either side of the interface.  Figure \ref{fig:FIXvsMOV.png} demonstrates the advantage of the moving mesh method in context of this work. Using a fixed mesh approach, the sharp compositional features can artificially diffuse as the ejecta expands over two to three orders of magnitude into logarithmically spaced computational zones.

The flux across the interface in Equation \ref{eq:timestep} is approximated via the HLLC Riemann solver \citep{Toro_1994}. The HLLC solver accounts for the existence of a contact discontinuity and pairs well with the moving mesh method to lower numerical diffusion, as long as we correctly choose the solution along the path traced out by the interface \citep{Duffell_2011}.

To achieve second order convergence, we use the Piecewise Linear Method (PLM) for spatial interpolation to the zone interfaces, and for higher order accuracy in time, we use a method of lines Runge-Kutta scheme described in \cite{Shu_1988}. However, PLM does not preserve the constraint that the sum of all mass fractions should be unity ($\sum_i  X_i = 1$). Consequently, \texttt{Cattails} renormalizes both the inner and outer PLM interpolated mass fractions at a given interface before passing the information to the Riemann solver. See \cite{Hasenour_2025} for a demonstration that this choice of numerical methods and coupling achieves second order convergence. 

\subsection{Gravity}
Following the methodology used in \texttt{RT1D} but detailed here first, \texttt{Cattails} includes self-gravity while explicitly conserving energy. This is achieved by reformulating the gravitational energy source term into a conservation law. First, the shell theorem greatly simplifies both the momentum and energy source terms $S^{(g)}$ in Equation \ref{eq:convo_laws} in spherically symmetric geometry as 
\begin{equation}
\label{eq:grav_source}
    \begin{gathered}
        S^{(g)}_{\epsilon} = \rho \mathbf{v} \cdot \mathbf{g} = -\rho v \frac{GM_{\rm enc}}{r^2} \\
        S^{(g)}_P = \rho g = -\rho \frac{GM_{\rm enc}}{r^2},
    \end{gathered}
\end{equation}
where $G$ is the gravitational constant and $M_{\rm enc}$ is the enclosed mass at a given radius $r$. If we now define an effective gravitational potential energy to be 
\begin{equation}
\label{eq:grav_energy}
    \epsilon_{g} \equiv -\rho \frac{GM_{\rm enc}}{r},
\end{equation}
then the energy source term in Equation \ref{eq:grav_source} can be expressed in terms of $\epsilon_g$. Specifically, it can be written as a derivative with respect to time and the divergence of a flux in the form
\begin{equation}
\label{eq:grav_convo}
    S^{(g)}_{\epsilon} = -\partial_t \epsilon_{g}+r^{-2}\partial_r (r^2\epsilon_{g} v).
\end{equation}

Treating gravity in this manner permits \texttt{Cattails} to include this effective gravitational potential energy and flux term into the total energy and explicitly conserve it during hydrodynamical steps. The gravitational momentum term remains a source term during the computations. 

Additionally, \texttt{Cattails} maintains hydrostatic equilibrium (HSE) to high precision in the unshocked regions of the white dwarf. While interpolating zone centered values to zone interfaces, we linearly reconstruct the pressure gradient needed to support gravity and subtract it off from the local pressures. \texttt{Cattails} performs a typical PLM step and then adds the HSE pressure component back to the solutions. This greatly reduces numerical errors in pressure balance for a self-gravitating system.

\subsection{Reactions}
In \texttt{Cattails}, the evolution of the composition during reaction steps is determined by the nuclear reaction network. The network used in this work is generated by \texttt{pynucastro} \citep{pynucastro2, pynucastro_zenodo}, and it contains 13 explicit isotopes: $^4$He, $^{12}$C, $^{16}$O, $^{20}$Ne, $^{24}$Mg, $^{28}$Si, $^{32}$S, $^{36}$Ar, $^{40}$Ca, $^{44}$Ti, $^{48}$Cr, $^{52}$Fe, and $^{56}$Ni. The processes of key interest are the triple alpha process, the alpha chain, heavy ion interactions, and approximations of $(\alpha,p)(p,\gamma)$ links through intermediate isotopes, as well as, reverse rates set by detailed balance. 

Numerically, the reaction network is commonly expressed in terms of molar abundances $Y_i = X_i/A_i$ (mass fraction over mass number) and time-evolution structured into an ordinary differential equation (ODE). The instantaneous derivatives across isotopes take the form
\begin{equation}
    \label{eq:rxn_ODE}
    \frac{d\mathbf{Y}}{dt}=\mathbf{f}(\rho, T, \mathbf{Y}),
\end{equation}
where $\mathbf{f}$ is referred to as the right-hand side (RHS) and contains all the rates affecting a given isotope for a given density $\rho$, temperature $T$, and composition $\mathbf{Y}$. \texttt{Cattails} progresses a state forward on hydrodynamical timescales using a variable order, semi-implicit, generalized step-doubling method ODE integrator with adaptive stepping. The underlying $4^{\rm th}$ order step is detailed in Appendix A of \cite{Hasenour_2025}. To reduce the computationally expensive calls to the RHS and corresponding Jacobian, \texttt{Cattails} attempts the first step of every implicit integration using a $2^{\rm nd}$ order method checking if burning is ongoing before switching to the full $4^{\rm th}$ order scheme. 

Each sub-step of a reaction network integration is performed under the assumption of hydrostatic burning, i.e. the density and temperature are held fixed. However, when burning is aggressive, the temperature is expected to vary significantly over one hydrodynamical timescale. \texttt{Cattails} accounts for this by updating the temperature after every successful sub-step assuming the binding energy change is transferred directly to thermal energy. This effectively evolves the temperature during the reaction step without explicitly integrating over the energy.  

\subsection{Timestep}
The global timestep in \texttt{Cattails} is set by the hydrodynamical evolution and determined by the Courant-Friedrichs-Lewy (CFL) condition common to numerical solutions of hyperbolic systems. The guiding principle is that information should only travel to directly neighboring zones over one timestep, which leads to the following condition on the length of the timestep, based on the sound speed:
\begin{equation}
\label{eq:cfl_condition}
    \Delta t = C_{\rm CFL}*\min_i \left[\frac{\Delta x_i}{c_s(\rho_i,T_i,\mathbf{X}_i)+|v_i-\bar{w}_i|}\right],
\end{equation}
where $\Delta t$ is the timestep, $\Delta x_i$ is the radial width of zone $i$, $c_s$ is the sound speed determined by the equation of state, $v_i$ is the local fluid velocity, and $\bar{w_i}$ is the average of the left and right interface velocities. The subtraction of the average mesh motion is a benefit of the moving mesh technique that allows for longer timesteps to be taken. We choose a $C_{\rm CFL}$ number of $0.2$. 

\section{White Dwarfs and Ignitions} \label{sec:initial_cond}

\begin{figure}[t]
\centering
\includegraphics[width=0.4625\textwidth]{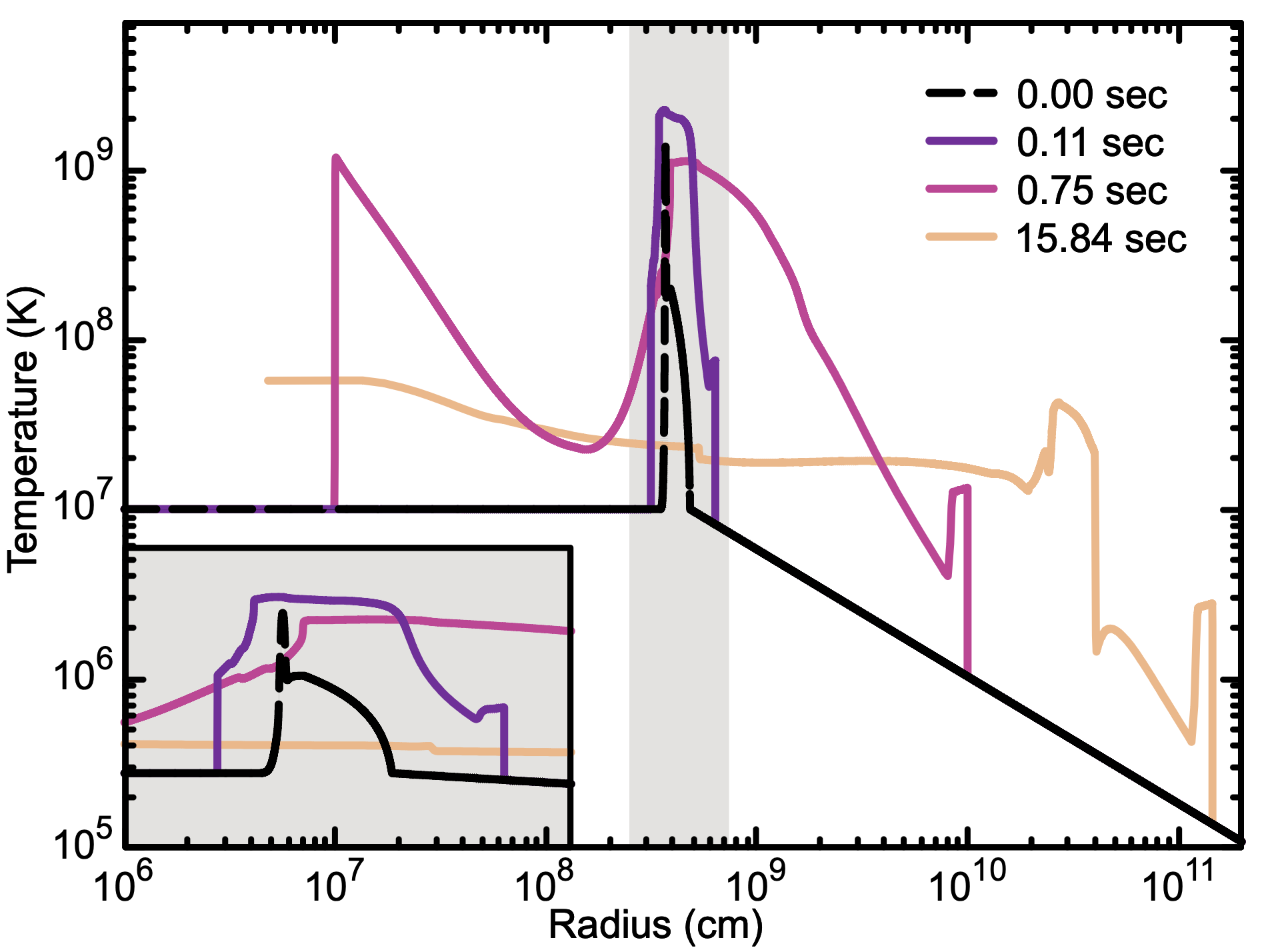}
\caption{The temperature profile evolution of the total mass \SI{1.1}{M_{\odot}} and shell mass \SI{0.05}{M_{\odot}} model. The initial profile (dashed black line) displays the isothermal core, isentropic shell, wind-like CSM, and ignition. The early time checkpoints (purple and magenta lines) capture both the inward and outward propagating burning fronts. The final snapshot at 15.84 seconds is long after core detonation and near homologous expansion.}
\label{fig:FDtemp.png}
\end{figure}

In this suite of double detonation models, we numerically construct each initial sub-Chandrasekhar mass WD in hydrostatic equilibrium following the prescription detailed in \cite{Zingale_2013}. Starting from the center, the highly degenerate 50/50 split C/O core is assumed to be isothermal at a temperature of \SI{e7}{K}. Next, a smooth transition region set by a hyperbolic tangent function of characteristic width \SI{5e6}{\cm} and spanning \SI{4e7}{\cm} brings the composition and temperature to the base of the helium shell values (pure helium at \SI{2e8}{K}). The shell proper is assumed to be isentropic, motivated by the expectation of the material to be convectively driven by slow burning at the base. Finally, we deviate from the \cite{Zingale_2013} prescription by switching to a power law in temperature $T \sim r^{-3/4}$ after the shell drops below \SI{e7}{K}. This is a computational choice so that once the material becomes radiation pressure dominated, the density falls off by a ``wind-like" profile $\rho \sim r^{-2}$. This helps us expand the domain out to \SI{2e12}{\cm} while keeping the total circumstellar material (CSM) mass negligible. Figure \ref{fig:FDtemp.png} provides an example of the initial temperature profile before and after ignition. 

In order to cover the large spatial domain, we distribute the initial zones using a hybrid function. Roughly 75$\%$ of the zones are linearly spaced out to \SI{e9}{\cm} ($\sim 2\times$ a WD radius), and the remaining 25$\%$ are distributed logarithmically out to the outer boundary. With a total of 32,768 computational zones, we reach a resolution of $\sim$\SI{0.4}{\km}, approaching the sub-kilometer resolution recommended by \cite{Brady_2025} for helium shell burning. The poorer resolution in zones with radius $r>$\SI{e9}{\cm} contributes to the large amount of numerical diffusion seen in the fixed mesh case in Figure \ref{fig:FIXvsMOV.png}. In the moving mesh case, the zones carrying the material expand with the fluid, and so drastically less numerical diffusion occurs while expanding into regions with much lower resolution.

\begin{figure}
\centering
\includegraphics[width=0.4625\textwidth]{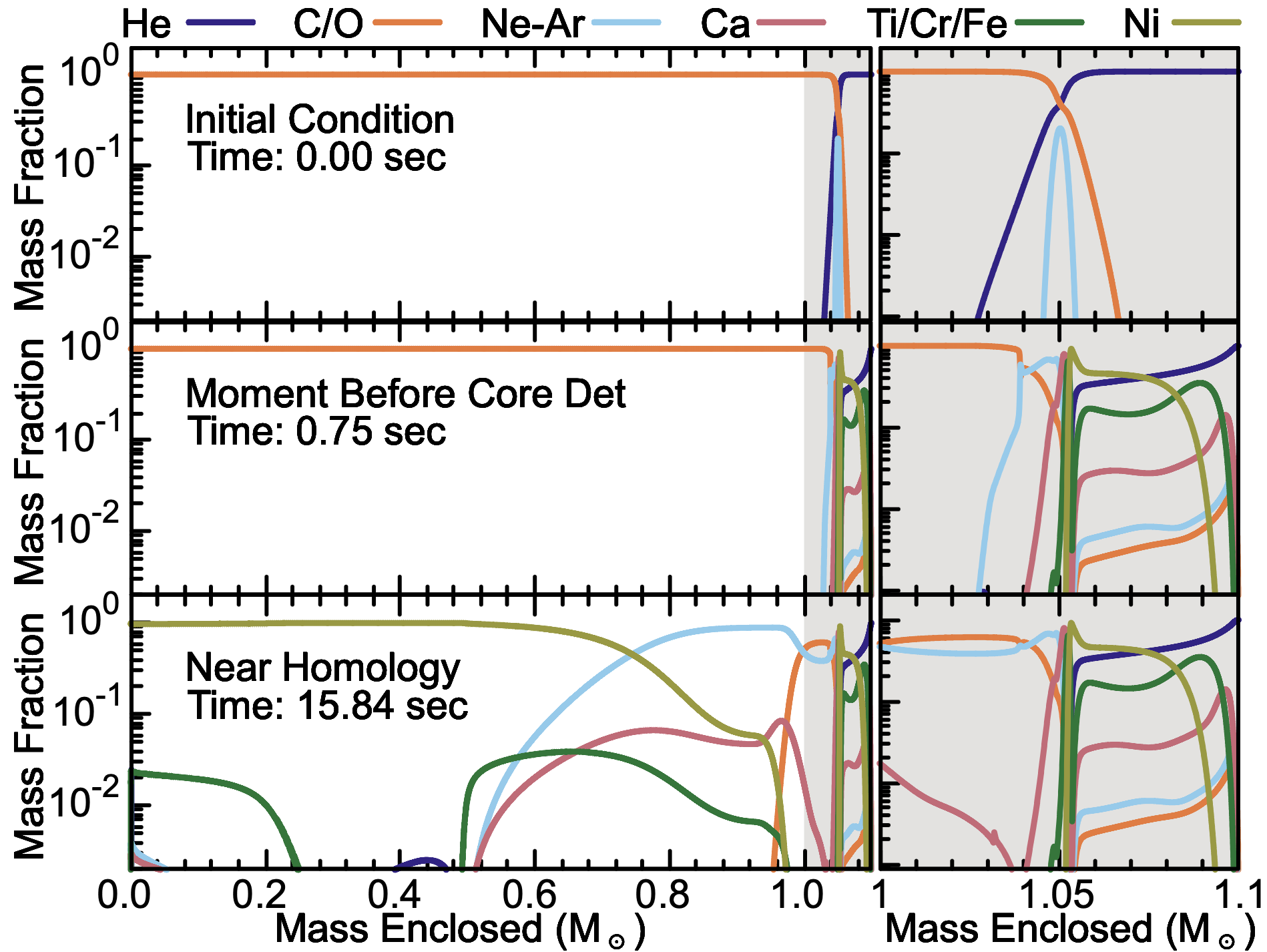}
\caption{Snapshots of the composition profiles for the total mass \SI{1.1}{M_{\odot}} and shell mass \SI{0.05}{M_{\odot}} model. (Top) The initial condition flips $20\%$ of the He, C, and O to Ne inside the transitional region to trigger thermonuclear runaway. (Middle) In the first detonation of this thick shell model, the helium shell burning produces Ni and other isotopes. (Bottom) The core detonation yields stratified Ni, Si/S, and O layers, surrounded by the helium shell ashes.}
\label{fig:FDpanels.png}
\end{figure}

The ignition method is novel. In 1D modeling, any change to the profile represents a change to the entire spherical shell at that radius. In an effort to introduce a small perturbation that leads to thermonuclear runaway, we focus on the most combustible material, that being the premixed material in the transition region between core and shell. Each double detonation model begins by flipping up to $20\%$ of the C/O/He to Ne in a gaussian profile of width \SI{1.17e6}{\cm}. The binding energy is transferred to the thermal energy, and a hot spot is formed with no added external energy. This ``flip detonation" is deemed successful when this region immediately begins runaway burning to IMEs and iron-group elements. These initial conditions in composition can be seen in the top panel of Figure \ref{fig:FDpanels.png}. In future studies, we intend to directly compare this ignition method against 2D and 3D models. 

\section{Models} 
\label{sec:models}

The suite of double detonation models is parametrized in terms of the initial white dwarf total mass (C/O core mass + He shell mass) and He shell mass covering four total masses: \SI{0.95}{M_{\odot}}, \SI{1.00}{M_{\odot}}, \SI{1.05}{M_{\odot}}, and \SI{1.10}{M_{\odot}}, and varying the shell mass across thin to thick shells from \SI{0.010}{M_{\odot}} -- \SI{0.080}{M_{\odot}}. This approach is motivated by one of the results of \cite{Polin_2019} where it was found that the total abundances are largely determined by the total mass of the initial white dwarf as this sets the density profile of the degenerate material. Table \ref{table:hydro} shows this relationship is reproduced well at higher total masses, but for the lower total masses, the strength of the converging shock from the helium detonation affects the final yields. 

\subsection{Nuclear Hydrodynamics Evolution}

A typical \texttt{Cattails} double detonation model proceeds as follows. In the first $\sim$\SI{0.01}{\second}, the ignition region at the base of the shell undergoes thermonuclear runaway and sends shocks radially inward and outward. The outward propagating front continues to burn the shell for another $\sim$\SI{0.1}{\second}, and the energy released unbinds most of the material. The atomic yields of the shell burning depend largely on the density of the material and the aggressiveness of the initial transitional material burning. Higher total masses and higher shell masses result in atomic yields of heavier ions also shown in Table \ref{table:hydro}. The thinnest shells only burn mild amounts of He to C. Intermediate shells can produce IMEs like Si, S, and Ca, with small amounts of radioactive Ti, Cr, and Fe. The thickest shells burn all throughout the alpha chain up to Ni. All models have an envelope of high velocity, unburnt helium unbound by the shock.

Meanwhile, the radially inward propagating front ceases burning but proceeds to the core hydrodynamically. After reaching a density profile shallower than $\rho \sim r^{-3}$, the converging shock begins to accelerate and strengthen. Once the C/O gets shocked to a temperature around \SI{2e9}{K}, the core ignites. The core detonation bounces off the inner reflecting boundary and proceeds, burning as it goes, throughout the white dwarf. By $\sim$\SI{10}{\second}, the ejecta is sufficiently close to homologous expansion. Figures \ref{fig:FDtemp.png} and \ref{fig:FDpanels.png} present snapshots of the evolution for the temperature profile and composition at moments of shell detonation, core detonation, and evolution end. The typical ejecta profile has a ball of Ni in the core. As the burning front proceeds to lower densities, the burning transitions to a Si/S dominated shell. Finally, the burning then transitions to producing primarily O before reaching the ashes of the shell. 

\begin{figure*}
\centering
\includegraphics[width=0.7\textwidth]{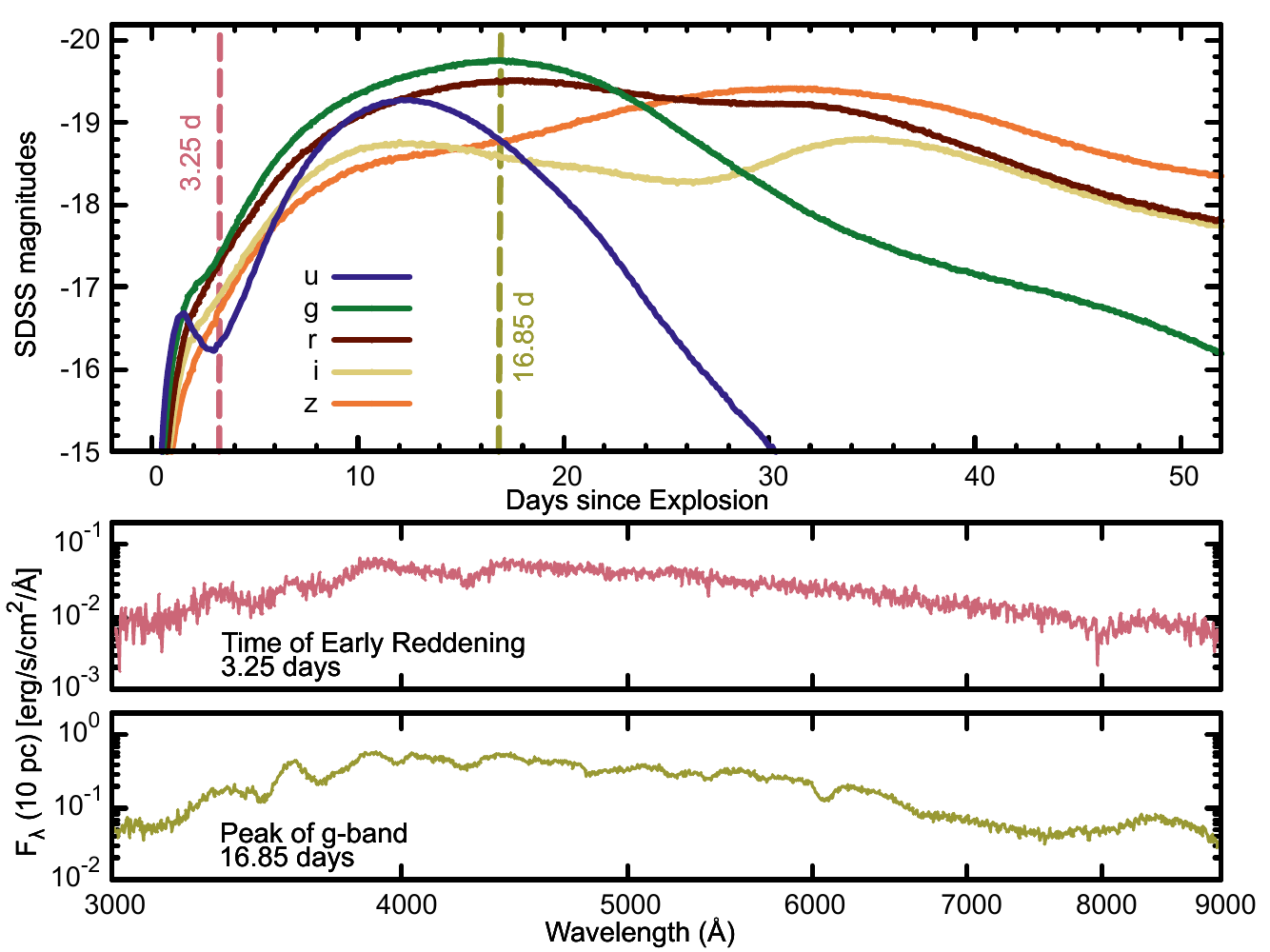}
\caption{(Top) The \textit{ugriz} light curve of the thick shell double detonation with total mass \SI{1.1}{M_{\odot}} and shell mass \SI{0.05}{M_{\odot}}. It shows an early bump at $\sim \SI{1.5}{d}$ from radioactive shell emission followed by reddening from line blanketing, both long before peak brightness. (Middle) The spectrum is provided at the checkpoint nearest the time of reddening ($\SI{3.25}{d}$), i.e. the time of the local maximum of the line blanketing effect. (Bottom) The spectrum nearest the g-band peak brightness ($\SI{16.85}{d}$) shows shallow absorption features expected of high total mass double detonation models.}
\label{fig:FD_lc_spec.pdf}
\end{figure*}

\subsection{Radiative Transfer Calculations}

The modeling of the supernovae emission from the double detonations into synthetic light curves and spectra is handled by the \texttt{Sedona} code \citep{Kasen_2006}. \texttt{Sedona} is a time-dependent, Monte Carlo radiative transfer code. In this suite of models, the calculations assume local thermodynamic equilibrium (LTE) to determine ionization and excitation fractions and homologous expansion for the ejecta. Three radioactive decay chains: $^{56}$Ni $\rightarrow$ $^{56}$Co $\rightarrow$ $^{56}$Fe, $^{52}$Fe $\rightarrow$ $^{52}$Mn $\rightarrow$ $^{52}$Cr, and $^{48}$Cr $\rightarrow$ $^{48}$V $\rightarrow$ $^{48}$Ti, generate additional luminosity that can eventually diffuse out to the observer. 

Each double detonation model from \texttt{Cattails} is mapped to \texttt{Sedona} by the following process. First, \texttt{Cattails} evolves the ejecta out to near-homologous expansion, typically after expanding by two to three orders of magnitude in radius. The characteristic time since explosion is then fit to the resulting velocity profile via a simple linear regression in log-log space. The hydrodynamical zones are then binned down to $\sim$500 zones conserving mass, momentum, and energy. Any lingering deviations from homology in velocity are then smoothed over by setting the velocity at a given radius to be the best-fit homologous velocity.

The time series of synthetic spectra produced in the radiative transfer calculations is sent through a collection of filters to generate light curves. In this study, we provide analysis in two filter systems: SDSS-2010 \textit{ugriz} in the AB magnitude system and Johnson/Cousins UBVRI in the Vega magnitude system. We separate the models into three qualitative behaviors (thin, intermediate, and thick). The light curves of thin shell models do not show any significant pre-peak early brightening or other sign of the helium shell ashes. The intermediate mass shells show reddening during the color evolution from line blanketing, but no light curve bump. The thickest/highest mass shells produce radioactive isotopes that decay and heat the ashes of the first detonation resulting in an early bump in light curves. This very early emission is immediately followed by reddening. Figure \ref{fig:FD_lc_spec.pdf} contains the light curve and two spectra of a high total mass, thick shell model showing both an early luminosity bump and significant reddening. Further, Appendix \ref{sec:rt_table}, Table \ref{table:rt_calc} summarizes the synthetic observables for all the models including the time and magnitude of peak brightness, as well as the time of the reddening and its corresponding color. 

\begin{figure*}
\centering
\includegraphics[width=0.4625\textwidth]{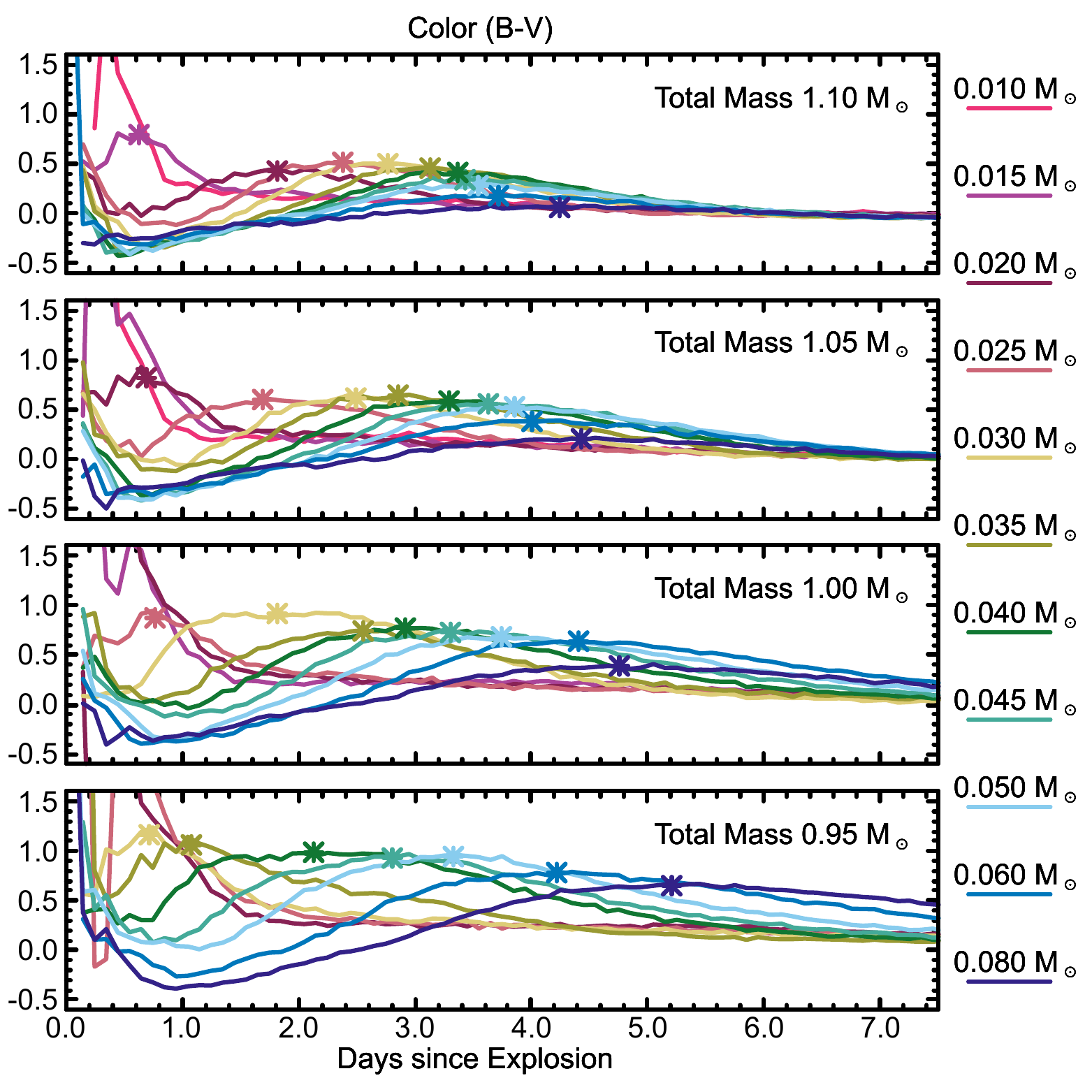}
\includegraphics[width=0.4625\textwidth]{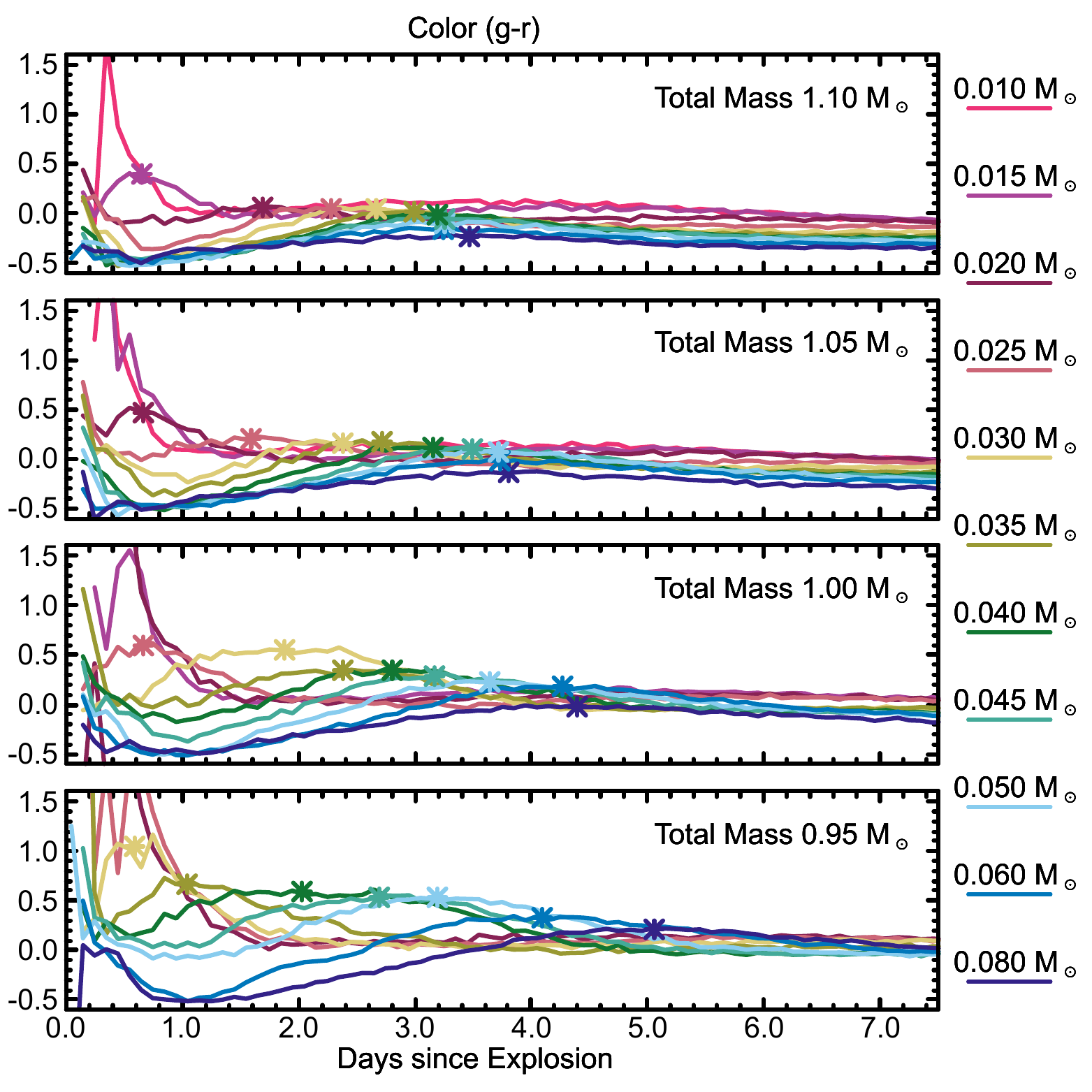}
\caption{Early time (Left) B-V color and (Right) g-r color evolution for all models in days since explosion. Thin shells recover the normal SNe Ia red to blue behavior. Intermediate shells have early time reddening. Thick shell models show early blue emission followed by reddening. Finally, the most massive shell models that produce significant Ni have the magnitude of reddening suppressed by the heating from decays.}
\label{fig:FD_color.png}
\end{figure*}

The focus of this work is on this early time reddening from the line blanketing of the helium shell ashes. Figure \ref{fig:FD_color.png} shows the color evolution in the first 10 days since explosion (long before the peak luminosity of the supernova). The duration of the reddening is characterized by the time at which the line blanketing effect reaches a local maximum. In practice, the feature is fit with a $5^{\rm th}$ order polynomial in a truncated region around the reddening. Likewise, the peak brightness is fit with a $5^{\rm th}$ order polynomial as well to obtain a precise measure of time since explosion. Table \ref{table:rt_calc} contains a summary of these fit observables for each model in the suite. For a given total mass, the most noticeable trend is that the width of the reddening and the time of its maximal effect increase with shell mass, as the photons have more material to diffuse through before escaping the ejecta. For the thicker shell models that produce radioactive isotopes, very blue emission is seen before the reddening and the heating from the decay in the shell pushes the magnitude of reddening down. 

\section{Reddening and Mass Relations} 
\label{sec:results}

The first order expectation for the relation between mass of the shell ($M_{\rm shell}$) and time of the reddening ($t_{\rm red}$) would be to evaluate the moment the photon diffusion timescale is comparable to the timescale of expansion of the ejecta \citep{Colgate_1969}. This criterion can be expressed in terms of the optical depth $\tau$: 

\begin{equation}
\label{eq:time_diffusion}
    \tau = \int \rho \kappa dr = \frac{c}{v} 
\end{equation}
where $\kappa$ is the opacity and $c$ is the speed of light. In the simple picture, the photons exhibiting the early reddening are produced at the photosphere at the base of the shell, and diffuse outward through the shell. Now assuming uniform values for the shell material, we can estimate the diffusive timescale above. Homologous expansion relates the characteristic radius and velocity of the shell, and together with Equation \ref{eq:time_diffusion}, this gives a relation for the time of the maximum early reddening in days since explosion $t_{\rm red}$ as

\begin{equation}
\label{eq:time_bump}
    t_{\rm red} \sim \sqrt{\frac{\kappa M_{\rm shell}}{(4\pi c)v_{\rm shell}}}.
\end{equation}

Using $\kappa = 0.2$ cm$^2$/g and choosing $v_{\rm shell}$ as the velocity measured at the base of the shell as approximate values, we arrive at the following estimate:

\begin{equation}
\label{eq:time_expect}
    t_{\rm red} \approx 10 {\rm d} \sqrt{\frac{M_{\rm shell}}{M_{\odot}}}.
\end{equation}

\begin{figure}[h]
\centering
\includegraphics[width=0.4625\textwidth]{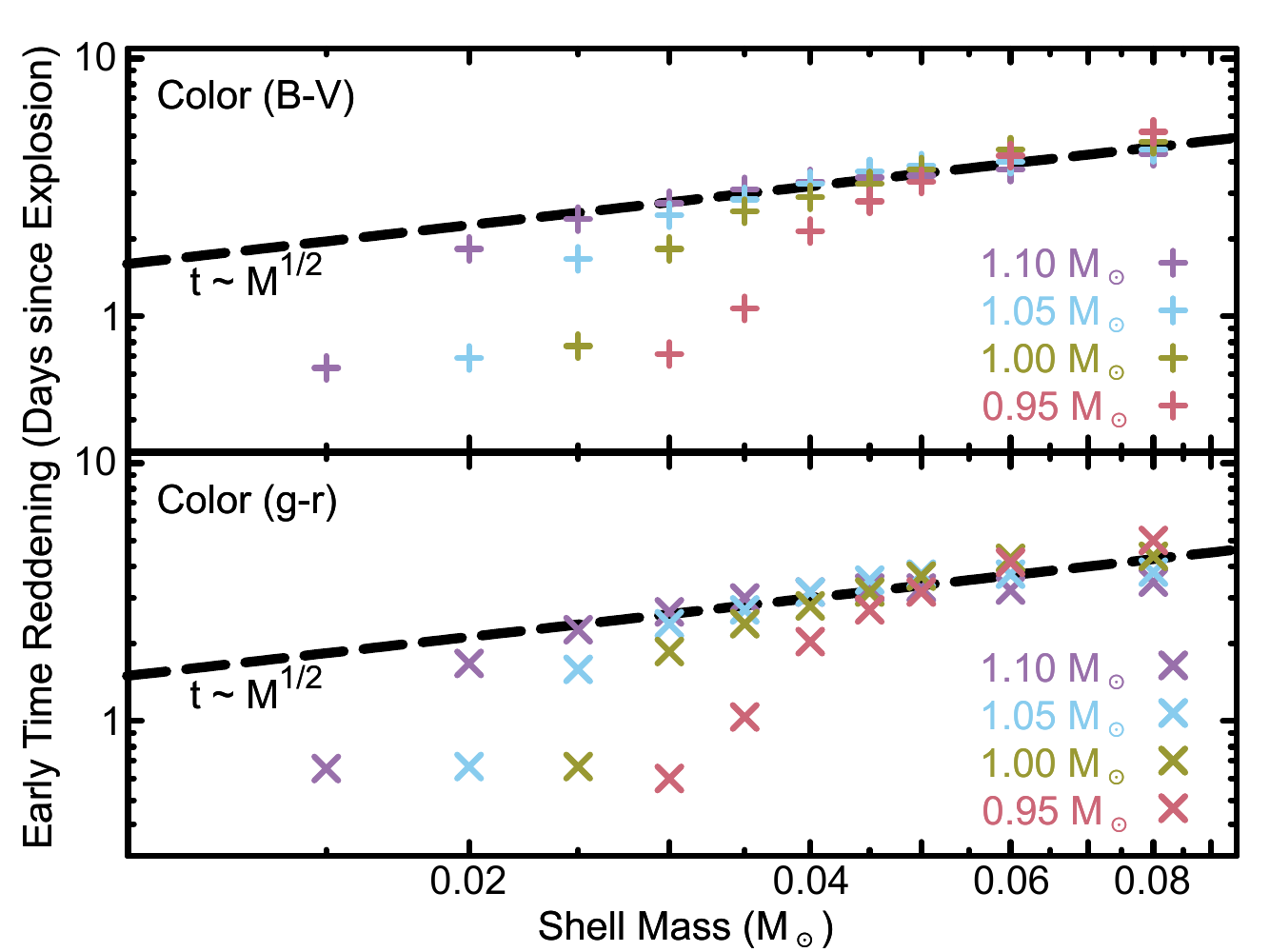}
\caption{Time of the early reddening maximum in days since explosion for all models in log-space. The black dashed line presents the estimate of Equation \ref{eq:time_expect} with proportionality constants 16 days and 15 days for B-V and g-r, respectively. Thicker shell models trend toward the estimated relation, while thinner shell models redden earlier than predicted.}
\label{fig:PEAKest.pdf}
\end{figure}

Comparing this expected relation to the numerical calculations in Figure \ref{fig:PEAKest.pdf}, we note that thicker shell models show rough agreement. But the thinner shells diverge from this trend and shift the reddening to earlier times. Further, this turn off is dependent on the total mass. To characterize this behavior, we modify Equation \ref{eq:time_expect} with a sharp exponential decay as

\begin{equation}
\label{eq:time_model}
    t_{\rm red}(M_{\rm shell}) = t_0 e^{-\left(\frac{m_0}{M_{\rm shell}}\right)^5}\sqrt{\frac{M_{\rm shell}}{M_{\odot}}}.
\end{equation}

\begin{figure}[h]
\centering
\includegraphics[width=0.4625\textwidth]{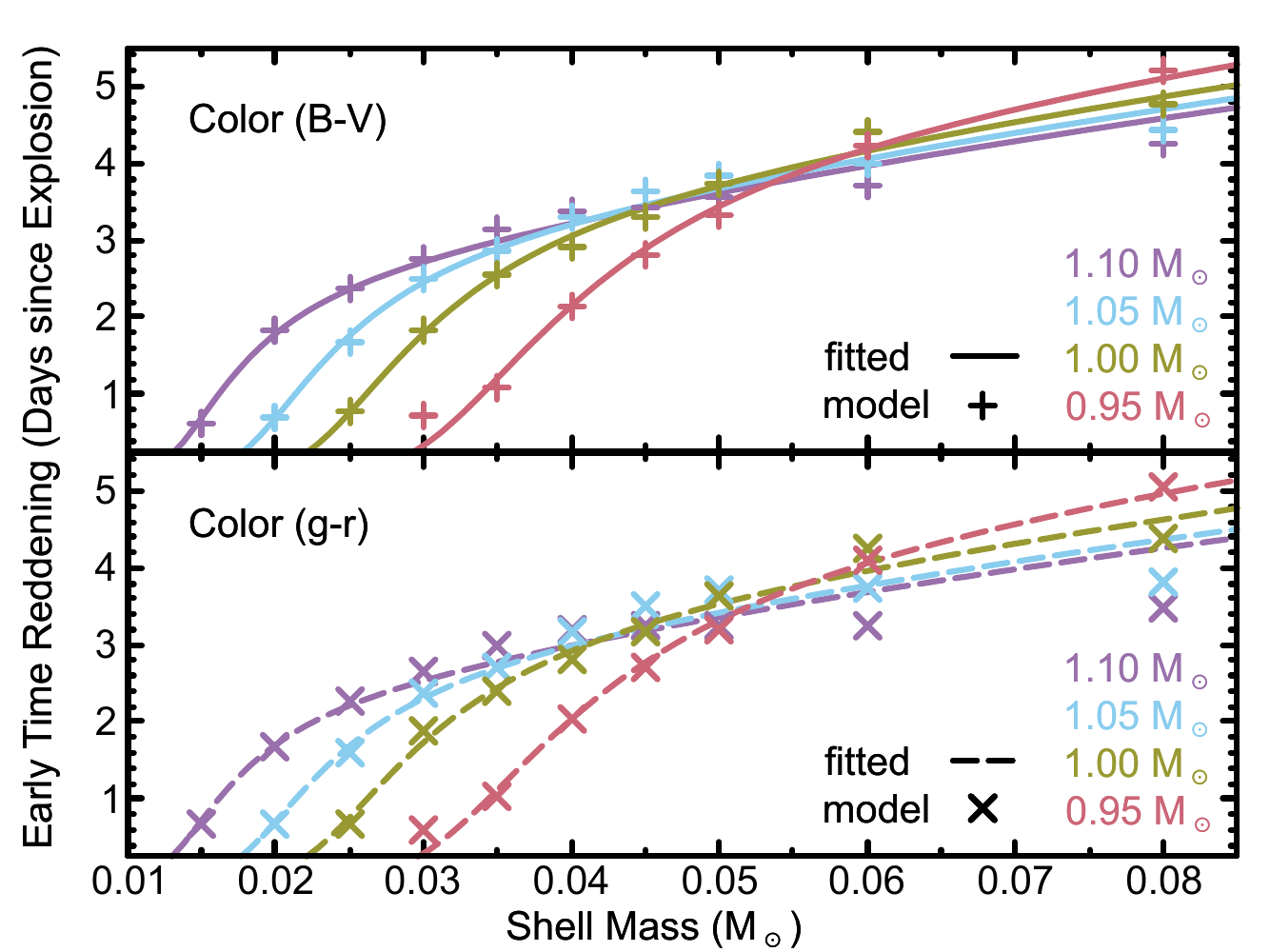}
\caption{The time of the early reddening maximum in days since explosion for all models in lin-space. The colored curves show the best fit relations as in Equation \ref{eq:time_model} and Table \ref{table:time_shell}.}
\label{fig:PEAKvsSHELL.pdf}
\end{figure}

For each total mass and color, we fit using a least squares method the two free parameters $t_0$ and $m_0$ across all total masses and filter systems. The results are plotted in Figure \ref{fig:PEAKvsSHELL.pdf} and summarized in Table \ref{table:time_shell}. Further, the parameters $t_0$ and $m_0$ show a predictable dependence on the total mass of the system, shifting the fitting function up/down and transitioning earlier/later. We found this dependence to be well fit by power laws of the form 

\begin{equation}
\label{eq:mtotal_dependence}
    \begin{gathered}
        t_0 (B-V) = 17.39d\left(\frac{M_{\rm tot}}{M_{\odot}}\right)^{-0.95}, \\
        m_0 (B-V) = 0.0265 M_{\odot} \left(\frac{M_{\rm tot}}{M_{\odot}}\right)^{-5.6}, \\
        t_0 (g-r) = 16.53d\left(\frac{M_{\rm tot}}{M_{\odot}}\right)^{-1.44}, \\
        m_0 (g-r) = 0.0264 M_{\odot} \left(\frac{M_{\rm tot}}{M_{\odot}}\right)^{-5.9}.
    \end{gathered}
\end{equation}

\begin{deluxetable}{||c|c|c|c||}
\setlength{\tabcolsep}{14pt}
\tablecaption{Reddening Best Fit Parameters\label{table:time_shell}}
\tablehead{
\colhead{Color} & \colhead{$M_{\rm tot}$} & \colhead{$t_0$} & \colhead{$m_0$}
}
\startdata
($B-V$) & 0.95 $M_\odot$ & 18.3 & 0.0353 \\
($B-V$) & 1.00 $M_\odot$ & 17.3 & 0.0263 \\
($B-V$) & 1.05 $M_\odot$ & 16.6 & 0.0209 \\
($B-V$) & 1.10 $M_\odot$ & 15.9 & 0.0150 \\
\hline
($g-r$) & 0.95 $M_\odot$ & 17.9 & 0.0357 \\
($g-r$) & 1.00 $M_\odot$ & 16.4 & 0.0261 \\
($g-r$) & 1.05 $M_\odot$ & 15.5 & 0.0205 \\
($g-r$) & 1.10 $M_\odot$ & 14.4 & 0.0145 \\
\enddata
\tablecomments{The relation between time of the early reddening maximum and initial shell mass is characterized by two parameters: $t_0$, the overall scale, and $m_0$, the transition to exponential decay. The parameters are fit for two colors and all total masses.}
\end{deluxetable}

\begin{figure}[h]
\centering
\includegraphics[width=0.4625\textwidth]{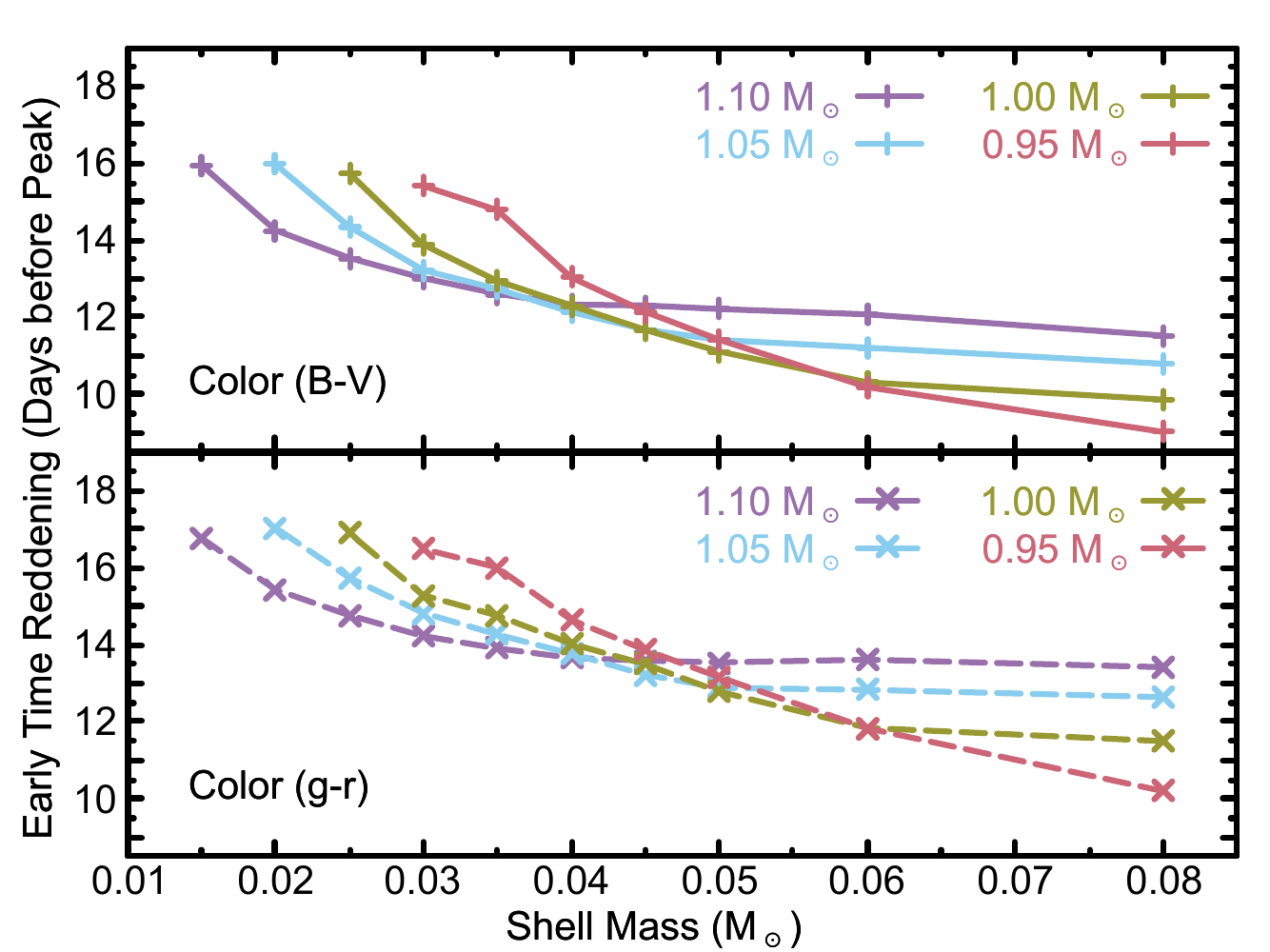}
\caption{Time of the early reddening maximum in days before peak brightness. Given a time of an observed reddening before peak, we can estimate the initial shell mass for different total masses. Very early reddening ($>$ 14 days before peak) are consistent with shell masses 0.02-0.04 $M_{\odot}$. Meanwhile, reddening closer to peak ($<$ 12 days before peak) only occur for models below 1.00 $M_{\odot}$ total mass with thick helium shells.}
\label{fig:PEAKb4PEAK.pdf}
\end{figure}

Now given an observed type Ia with an early reddening, we can ask what shell mass is consistent with the time of the feature. However, the above characterization is difficult to apply to observations because it is in terms of time since explosion, which is difficult to determine for a real event. Figure \ref{fig:PEAKb4PEAK.pdf} instead plots the time of the reddening in days before peak brightness. The estimation using the photon diffusion timescale no longer holds as the time of peak brightness itself depends on total mass and shell mass. Given a total mass, thicker shells will lead to earlier peaks as stronger converging shocks will result in more energetic explosions. The higher kinetic energies expand the material quicker and lower opacities. Table \ref{table:rt_calc} shows this relation with the exception of the thickest shell models that produce significant Ni in the shell. The heating from the radioactive decay drives emission to be bluer and competes with the reddening from line blanketing.

Figure \ref{fig:PEAKb4PEAK.pdf} can be used as an estimation tool for observations. Given a early reddening and peak, we can see the spread of shell masses/total masses consistent with the time of the event. The early reddening around 15 days before peak are consistent with different combinations of shell mass and total mass, but the later reddening around 10 days are only consistent with lower total mass models. However, this tool becomes significantly more powerful given an estimate for the total mass. Using the relation in Figure \ref{fig:PEAKb4PEAK.pdf}, an estimate of the total mass and time of the reddening directly gives an estimate for the initial shell mass, and so it is valuable to also estimate the total mass from a light curve.

\begin{figure}[h]
\centering
\includegraphics[width=0.4625\textwidth]{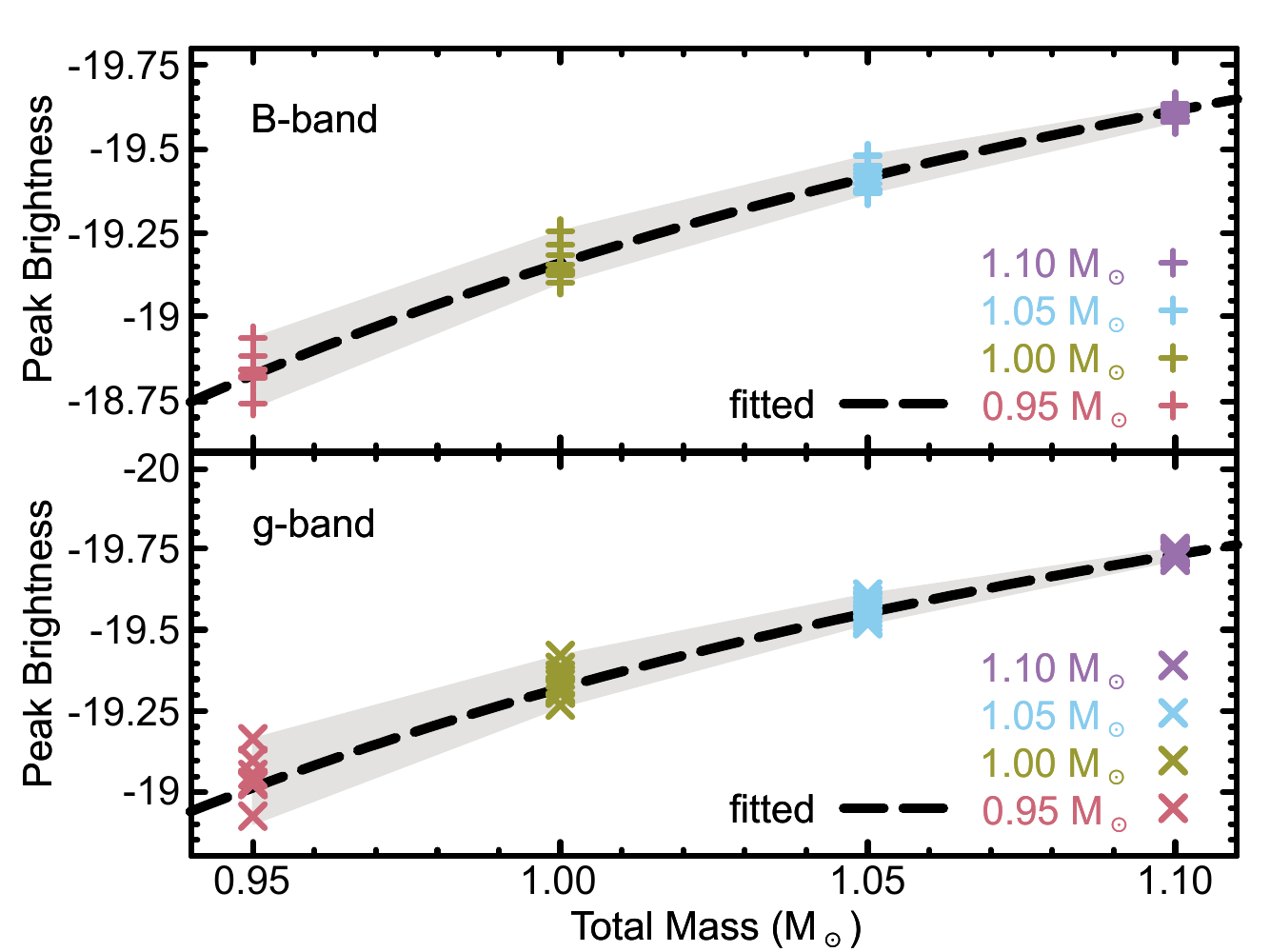}
\caption{Peak brightness in B-band and g-band for all models. Higher total mass models result in higher radioactive yields and a brighter supernovae with very little dependence on the shell mass. Lower total mass models have a greater dependence on the initial shell mass as the strength of the converging shock depends on the aggressiveness of the helium burning.}
\label{fig:PEAKbrightness.pdf}
\end{figure}

In addition to the time of peak brightness, we can also relate the magnitude of peak brightness to the masses of the initial WD of the models. Figure \ref{fig:PEAKbrightness.pdf} shows the trend that higher total mass models result in brighter supernovae. A sense of spread is given by the different shell masses, where lower total masses depend more significantly on the strength of the converging shock. We provide a fitting function for this relation that can be used to estimate the total mass of an observed supernovae given its peak brightness taking the following form

\begin{equation}
\label{eq:peak_brightness}
    \begin{gathered}
        B_{\rm mag} = -20.5+1.34\left(\frac{M_{\rm tot}}{M_{\rm \odot}}\right)^{-4.33},\\
        g_{\rm mag} = -20.5+1.18\left(\frac{M_{\rm tot}}{M_{\rm \odot}}\right)^{-4.52}.
    \end{gathered}
\end{equation}
$B_{\rm mag}$ and $g_{\rm mag}$ are the absolute magnitudes in B-band and g-band at peak brightness. Together the relations in Equations \ref{eq:time_model}, \ref{eq:mtotal_dependence}, and \ref{eq:peak_brightness} (or equivalently Figures \ref{fig:PEAKb4PEAK.pdf} and \ref{fig:PEAKbrightness.pdf}) provide a tool for recovering progenitor information from an observed light curve. 

\section{Discussion} \label{sec:discussion}

In this work, we introduce the \texttt{Cattails} code, a moving-mesh nuclear hydrodynamics code applicable to the of modeling thermonuclear supernovae. It is publicly available on \texttt{Github} at \url{https://github.com/DHasenour/Cattails} where we will continue development. In addition, a frozen version used in the this work's computational modeling will be available on \texttt{Zenodo} at \cite{Cattails_suite}.

\texttt{Cattails} is then used to compute a numerical suite of SNe Ia models in the style of a sub-Chandrasekhar double detonation. The models are ignited by flipping a small percentage of the composition up the alpha-chain, conserving the total energy, and evolved out the near homologous expansion. Synthetic light curves and spectra are then calculated using the radiative transfer code \texttt{Sedona}. Both the hydrodynamical outputs and the synthetic observables for all models will also be made available at \cite{Cattails_suite}. 

The atomic yields of the detonation models are tabulated in terms of total WD material and shell WD material. We recover the result in \cite{Polin_2019} that the total yields are largely determined by the initial total mass. We expand upon this by showing that the dependence on the shell mass for both the total yields and peak brightness grows and becomes significant for lower total mass models. Meanwhile, the shell yields greatly depend on the both the initial total mass and shell mass. Burning in thinner shells at lower densities can result in mild C production, intermediate shells produce IMEs often dominated by Si, S, and Ca, and thick shells yield significant radioactive isotope mass fractions including Ti, Cr, Fe, and Ni. 

In the synthetic observables, we focus on and fit the early time reddening maximum (if present) and the peak brightness of each model. First, we estimate a relation that the time of the reddening should scale by the shell as $t_{\rm red} \sim \sqrt{M_{\rm shell}}$, and then we demonstrate that intermediate and thick shells follow this expectation. Thinner shell models deviate from the analytical expectation and redden earlier, so we make an empirical relationship to capture this trend. Further, this relation shows dependence on the total mass shallowly in overall scale $t_0$, but steeply in turn off mass $m_0$. Meanwhile, the magnitude of peak brightness also depends on total mass with higher total masses resulting in brighter events. Inverted, this relation can estimate the total mass of the progenitor, and together with the relation of the reddening and shell mass, we form a series of tools to estimate the size of the shell that produced an observed early reddening in a SNe Ia. 

It should be emphasized that using these tools to interpret observations will only provide estimates for the total mass and shell mass consistent with the supernova; the amount of error in these estimates is not known. Multidimensional studies have shown viewing-angle dependence of peak magnitude, and it is possible that $t_{\rm red}$ also depends on viewing angle \citep{Collins_2025, Boos_2025, Callan_2026}. Non-LTE effects might matter; although the reddening occurs at very early times, it also arrives from the outermost material, which becomes optically thin very early. Typically, the LTE assumption is a reasonable approximation at early times because the bulk of the ejecta is optically thick and the photons interact regularly with the local material, but we have not tested whether this is affects the result. Also, these results might be impacted by the size of the reaction network used. Nevertheless, we have provided a framework of estimates and fitting functions that can be improved upon using more detailed numerical models, and we expect the qualitative trends to remain even if the fitted parameters get adjusted with more physics. This study will be the building block of tools to come back to in future studies as more physics is added.

\begin{acknowledgments}
Numerical calculations were preformed on the Petunia computing cluster hosted by the Department of Physics and Astronomy at Purdue University. We thank Chris J. Orr for his extensive help with maintaining Petunia. 

The self-gravity module of RT1D/Cattails was developed several years ago out of long conversations and long coding sessions between P.~D. and Bill Paxton.  We wish to acknowledge Bill's indelible impact on the field of astrophysics.
\end{acknowledgments}

%

\vspace{5mm}





\appendix
\section{Atomic Yields} \label{sec:hydro_table}

The nuclear yields produced in the \texttt{Cattails} double detonation models are presented in Table \ref{table:hydro}. The first four columns summarize the initial parameters including total mass and shell mass of the white dwarf and the radius and density were the ignition is centered. The next two sets of columns detail the nuclear yields of the shell material and the total material, respectively, in four groups: the light nuclei ($^4$He, $^{12}$C, $^{16}$O, and $^{20}$Ne), the IMEs ($^{24}$Mg, $^{28}$Si, $^{32}$S, $^{36}$Ar, and $^{40}$Ca), the non-nickel radioactive isotopes ($^{44}$Ti, $^{48}$Cr, and $^{52}$Fe), and lastly $^{56}$Ni.

\begin{deluxetable*}{||cccc|cccc|cccc||}
\setlength{\tabcolsep}{4pt}
\tablecaption{Hydrodynamical initial parameters and nucleosynthetic yields for all models.\label{table:hydro}}
\tablehead{
\multicolumn{2}{c}{Initial} & \colhead{$R_{\rm ign}$} & \colhead{$\rho_{\rm ign}$} & \multicolumn{4}{c}{Shell Yields} & \multicolumn{4}{c}{Total Yields} \\
\hline
\colhead{$M_{\rm tot}$} & \colhead{$M_{\rm shell}$} & \colhead{$\times10^8$} & \colhead{$\times10^5$} & \colhead{He~-~Ne} & \colhead{Mg~-~Ca} & \colhead{Ti~-~Fe} & \colhead{Ni} & \colhead{He~-~Ne} & \colhead{Mg~-~Ca} & \colhead{Ti~-~Fe} & \colhead{Ni} \\
\colhead{(\unit{M_\odot})} & \colhead{(\unit{M_\odot})} & \colhead{(\unit{\cm})} & \colhead{(\unit{\gram \per \cm^3})} & \colhead{($M_\odot$)} & \colhead{($M_\odot$)} & \colhead{($M_\odot$)} & \colhead{($M_\odot$)} & \colhead{($M_\odot$)} & \colhead{($M_\odot$)} & \colhead{($M_\odot$)} & \colhead{($M_\odot$)}
}
\startdata
0.95 & 0.020 & 4.98 & 3.17 & 1.8e-02 & 2.0e-03 & 2.4e-06 & 1.2e-16 & 1.1e-01 & 4.9e-01 & 1.6e-02 & 3.3e-01 \\
0.95 & 0.025 & 4.91 & 3.80 & 2.2e-02 & 2.6e-03 & 3.7e-05 & 1.0e-12 & 1.2e-01 & 4.8e-01 & 1.6e-02 & 3.4e-01 \\
0.95 & 0.030 & 4.84 & 4.41 & 2.7e-02 & 3.3e-03 & 1.7e-04 & 2.9e-10 & 1.2e-01 & 4.7e-01 & 1.5e-02 & 3.5e-01 \\
0.95 & 0.035 & 4.78 & 5.00 & 2.3e-02 & 1.1e-02 & 5.1e-04 & 2.2e-08 & 1.1e-01 & 4.5e-01 & 1.5e-02 & 3.8e-01 \\
0.95 & 0.040 & 4.72 & 5.58 & 2.7e-02 & 1.0e-02 & 2.3e-03 & 7.2e-07 & 1.2e-01 & 4.4e-01 & 1.7e-02 & 3.7e-01 \\
0.95 & 0.045 & 4.67 & 6.15 & 3.0e-02 & 1.2e-02 & 3.7e-03 & 9.9e-06 & 1.2e-01 & 4.4e-01 & 1.8e-02 & 3.8e-01 \\
0.95 & 0.050 & 4.62 & 6.72 & 3.1e-02 & 1.2e-02 & 6.5e-03 & 6.9e-05 & 1.2e-01 & 4.3e-01 & 2.1e-02 & 3.8e-01 \\
0.95 & 0.060 & 4.53 & 7.83 & 3.4e-02 & 1.0e-02 & 1.5e-02 & 6.4e-04 & 1.2e-01 & 4.1e-01 & 2.8e-02 & 3.9e-01 \\
0.95 & 0.080 & 4.38 & 10.02 & 4.0e-02 & 8.1e-03 & 2.3e-02 & 8.7e-03 & 1.1e-01 & 3.9e-01 & 3.6e-02 & 4.1e-01 \\
\hline
1.00 & 0.015 & 4.74 & 3.15 & 1.3e-02 & 1.6e-03 & 1.6e-06 & 5.8e-17 & 8.2e-02 & 4.3e-01 & 1.6e-02 & 4.7e-01 \\
1.00 & 0.020 & 4.66 & 3.97 & 1.8e-02 & 2.3e-03 & 5.0e-05 & 3.9e-12 & 8.7e-02 & 4.3e-01 & 1.6e-02 & 4.7e-01 \\
1.00 & 0.025 & 4.59 & 4.74 & 2.1e-02 & 3.3e-03 & 2.7e-04 & 2.6e-09 & 9.1e-02 & 4.2e-01 & 1.6e-02 & 4.8e-01 \\
1.00 & 0.030 & 4.52 & 5.50 & 2.0e-02 & 7.9e-03 & 2.1e-03 & 5.4e-07 & 8.9e-02 & 4.1e-01 & 1.7e-02 & 4.9e-01 \\
1.00 & 0.035 & 4.47 & 6.23 & 2.3e-02 & 9.1e-03 & 2.8e-03 & 1.0e-05 & 9.1e-02 & 4.0e-01 & 1.8e-02 & 4.9e-01 \\
1.00 & 0.040 & 4.41 & 6.95 & 2.5e-02 & 9.5e-03 & 5.4e-03 & 1.0e-04 & 9.2e-02 & 3.9e-01 & 2.0e-02 & 4.9e-01 \\
1.00 & 0.045 & 4.36 & 7.66 & 2.7e-02 & 8.9e-03 & 9.1e-03 & 4.1e-04 & 9.3e-02 & 3.9e-01 & 2.4e-02 & 5.0e-01 \\
1.00 & 0.050 & 4.32 & 8.36 & 2.8e-02 & 8.0e-03 & 1.3e-02 & 1.0e-03 & 9.3e-02 & 3.8e-01 & 2.7e-02 & 5.0e-01 \\
1.00 & 0.060 & 4.24 & 9.75 & 3.1e-02 & 7.0e-03 & 1.8e-02 & 4.1e-03 & 9.3e-02 & 3.7e-01 & 3.2e-02 & 5.1e-01 \\
1.00 & 0.080 & 4.09 & 12.48 & 3.6e-02 & 6.0e-03 & 1.6e-02 & 2.3e-02 & 9.3e-02 & 3.5e-01 & 2.9e-02 & 5.3e-01 \\
\hline
1.05 & 0.010 & 4.50 & 2.90 & 8.9e-03 & 1.1e-03 & 2.5e-07 & 7.5e-19 & 5.8e-02 & 3.7e-01 & 1.7e-02 & 6.0e-01 \\
1.05 & 0.015 & 4.41 & 3.99 & 1.3e-02 & 1.9e-03 & 4.4e-05 & 3.5e-12 & 6.3e-02 & 3.7e-01 & 1.7e-02 & 6.0e-01 \\
1.05 & 0.020 & 4.33 & 5.01 & 1.7e-02 & 3.1e-03 & 3.5e-04 & 1.2e-08 & 6.7e-02 & 3.6e-01 & 1.7e-02 & 6.0e-01 \\
1.05 & 0.025 & 4.26 & 5.99 & 1.7e-02 & 6.4e-03 & 1.6e-03 & 2.7e-06 & 6.8e-02 & 3.6e-01 & 1.7e-02 & 6.1e-01 \\
1.05 & 0.030 & 4.20 & 6.93 & 1.9e-02 & 7.3e-03 & 3.6e-03 & 7.4e-05 & 7.0e-02 & 3.5e-01 & 1.9e-02 & 6.1e-01 \\
1.05 & 0.035 & 4.15 & 7.87 & 2.1e-02 & 7.0e-03 & 6.8e-03 & 4.4e-04 & 7.1e-02 & 3.4e-01 & 2.2e-02 & 6.1e-01 \\
1.05 & 0.040 & 4.10 & 8.77 & 2.2e-02 & 6.3e-03 & 1.0e-02 & 1.2e-03 & 7.2e-02 & 3.4e-01 & 2.5e-02 & 6.1e-01 \\
1.05 & 0.045 & 4.05 & 9.67 & 2.4e-02 & 5.9e-03 & 1.3e-02 & 2.5e-03 & 7.3e-02 & 3.3e-01 & 2.8e-02 & 6.2e-01 \\
1.05 & 0.050 & 4.01 & 10.56 & 2.5e-02 & 5.6e-03 & 1.4e-02 & 5.2e-03 & 7.4e-02 & 3.3e-01 & 2.9e-02 & 6.2e-01 \\
1.05 & 0.060 & 3.93 & 12.31 & 2.8e-02 & 5.1e-03 & 1.3e-02 & 1.4e-02 & 7.5e-02 & 3.2e-01 & 2.7e-02 & 6.3e-01 \\
1.05 & 0.080 & 3.80 & 15.78 & 3.2e-02 & 4.8e-03 & 1.0e-02 & 3.3e-02 & 7.5e-02 & 3.0e-01 & 2.4e-02 & 6.5e-01 \\
\hline
1.10 & 0.010 & 4.15 & 3.74 & 8.6e-03 & 1.4e-03 & 1.6e-05 & 1.5e-13 & 4.3e-02 & 3.1e-01 & 1.7e-02 & 7.3e-01 \\
1.10 & 0.015 & 4.06 & 5.17 & 1.2e-02 & 2.5e-03 & 3.4e-04 & 1.8e-08 & 4.8e-02 & 3.0e-01 & 1.7e-02 & 7.3e-01 \\
1.10 & 0.020 & 3.98 & 6.45 & 1.3e-02 & 5.0e-03 & 1.7e-03 & 1.2e-05 & 5.0e-02 & 3.0e-01 & 1.8e-02 & 7.3e-01 \\
1.10 & 0.025 & 3.92 & 7.70 & 1.5e-02 & 5.5e-03 & 4.0e-03 & 2.7e-04 & 5.2e-02 & 3.0e-01 & 2.0e-02 & 7.3e-01 \\
1.10 & 0.030 & 3.87 & 8.92 & 1.7e-02 & 5.0e-03 & 7.0e-03 & 1.0e-03 & 5.4e-02 & 2.9e-01 & 2.3e-02 & 7.3e-01 \\
1.10 & 0.035 & 3.82 & 10.12 & 1.8e-02 & 4.6e-03 & 9.6e-03 & 2.4e-03 & 5.5e-02 & 2.9e-01 & 2.5e-02 & 7.3e-01 \\
1.10 & 0.040 & 3.77 & 11.29 & 2.0e-02 & 4.4e-03 & 1.1e-02 & 5.2e-03 & 5.7e-02 & 2.9e-01 & 2.6e-02 & 7.3e-01 \\
1.10 & 0.045 & 3.73 & 12.45 & 2.1e-02 & 4.3e-03 & 1.0e-02 & 9.6e-03 & 5.8e-02 & 2.8e-01 & 2.5e-02 & 7.4e-01 \\
1.10 & 0.050 & 3.69 & 13.59 & 2.2e-02 & 4.2e-03 & 9.1e-03 & 1.4e-02 & 5.8e-02 & 2.8e-01 & 2.4e-02 & 7.4e-01 \\
1.10 & 0.060 & 3.62 & 15.88 & 2.5e-02 & 4.1e-03 & 8.0e-03 & 2.3e-02 & 5.9e-02 & 2.7e-01 & 2.3e-02 & 7.5e-01 \\
1.10 & 0.080 & 3.49 & 20.39 & 2.8e-02 & 4.1e-03 & 6.5e-03 & 4.1e-02 & 6.1e-02 & 2.6e-01 & 2.1e-02 & 7.6e-01 \\
\enddata
\tablecomments{The hydrodynamical initial parameters are described by the total mass of the WD, the shell mass of the WD, and the radius and density at which the ignition is placed. The final atomic yields are grouped into radioactive Ni, other radioactive isotopes (Ti-Fe), IMEs (Mg-Ca), and light elements (He-Ne). The total yields are largely determined by the total mass, but the shell yields depend strongly on both the total mass and shell mass.}
\end{deluxetable*}

\section{Synthetic Observables} \label{sec:rt_table}

The results of the \texttt{Sedona} radiative transfer calculations are summarized in Table \ref{table:rt_calc}. The first column again parametrizes the initial white dwarf total and shell mass. The next set of columns lists the bessell filter system (\textit{UBVRI}) values  at peak brightness for time since explosion and B-band absolute magnitude. For models that have a significant red color bump, the next columns detail fit values for days since explosion and the local maximum color. A blank entry here means that this double detonation is a thin shell model and the helium shell ash does not substantially affect the light curve. The remaining columns repeat this for the SDSS-2010 (\textit{ugriz}) filter system in the AB magnitude system. 

\begin{deluxetable*}{||cc|cc|cc|cc|cc||}
\tablecaption{Synthetic observables for all models. \label{table:rt_calc}}
\tablehead{
\multicolumn{2}{c}{Initial} & \multicolumn{2}{c}{Peak Brightness ($B$)} & \multicolumn{2}{c}{Reddening ($B-V$)} & \multicolumn{2}{c}{Peak Brightness ($g$)} & \multicolumn{2}{c}{Reddening ($g-r$)} \\
\cline{1-2} \cline{3-4} \cline{5-6} \cline{7-8} \cline{9-10}
\colhead{$M_{\rm tot}$} & \colhead{$M_{\rm shell}$} & \colhead{$t$} & \colhead{$B$} & \colhead{$t$} & \colhead{$B-V$} & \colhead{$t$} & \colhead{$g$} & \colhead{$t$} & \colhead{$g-r$} \\
\colhead{($M_\odot$)} & \colhead{($M_\odot$)} & \colhead{(days)} & \colhead{(mag)} & \colhead{(days)} & \colhead{(mag)} & \colhead{(days)} & \colhead{(mag)} & \colhead{(days)} & \colhead{(mag)}
}
\startdata
0.95 & 0.020 & 16.48 & $-18.74$ & \nodata & \nodata & 17.19 & $-18.90$ & \nodata & \nodata \\
0.95 & 0.025 & 16.31 & $-18.74$ & \nodata & \nodata & 17.15 & $-18.91$ & \nodata & \nodata \\
0.95 & 0.030 & 16.13 & $-18.75$ & 0.72 & 1.16 & 17.08 & $-18.93$ & 0.60 & 1.04 \\
0.95 & 0.035 & 15.86 & $-18.85$ & 1.08 & 1.05 & 17.05 & $-19.02$ & 1.04 & 0.67 \\
0.95 & 0.040 & 15.16 & $-18.82$ & 2.13 & 0.99 & 16.68 & $-19.02$ & 2.03 & 0.59 \\
0.95 & 0.045 & 14.95 & $-18.83$ & 2.81 & 0.93 & 16.56 & $-19.03$ & 2.70 & 0.54 \\
0.95 & 0.050 & 14.74 & $-18.84$ & 3.33 & 0.95 & 16.36 & $-19.05$ & 3.20 & 0.53 \\
0.95 & 0.060 & 14.40 & $-18.89$ & 4.22 & 0.78 & 15.93 & $-19.10$ & 4.09 & 0.33 \\
0.95 & 0.080 & 14.22 & $-18.94$ & 5.20 & 0.66 & 15.25 & $-19.16$ & 5.06 & 0.21 \\
\hline
1.00 & 0.015 & 16.95 & $-19.11$ & \nodata & \nodata & 17.75 & $-19.26$ & \nodata & \nodata \\
1.00 & 0.020 & 16.71 & $-19.10$ & \nodata & \nodata & 17.69 & $-19.26$ & \nodata & \nodata \\
1.00 & 0.025 & 16.51 & $-19.10$ & 0.77 & 0.87 & 17.57 & $-19.27$ & 0.67 & 0.59 \\
1.00 & 0.030 & 15.71 & $-19.14$ & 1.82 & 0.92 & 17.17 & $-19.31$ & 1.88 & 0.56 \\
1.00 & 0.035 & 15.50 & $-19.13$ & 2.55 & 0.74 & 17.15 & $-19.30$ & 2.39 & 0.34 \\
1.00 & 0.040 & 15.23 & $-19.14$ & 2.92 & 0.76 & 16.83 & $-19.32$ & 2.81 & 0.35 \\
1.00 & 0.045 & 14.98 & $-19.16$ & 3.31 & 0.73 & 16.67 & $-19.34$ & 3.17 & 0.28 \\
1.00 & 0.050 & 14.86 & $-19.18$ & 3.75 & 0.68 & 16.43 & $-19.36$ & 3.64 & 0.23 \\
1.00 & 0.060 & 14.71 & $-19.21$ & 4.40 & 0.64 & 16.11 & $-19.38$ & 4.26 & 0.18 \\
1.00 & 0.080 & 14.62 & $-19.26$ & 4.76 & 0.39 & 15.89 & $-19.42$ & 4.39 & $-0.02$ \\
\hline
1.05 & 0.010 & 17.20 & $-19.39$ & \nodata & \nodata & 17.91 & $-19.53$ & \nodata & \nodata \\
1.05 & 0.015 & 16.96 & $-19.38$ & \nodata & \nodata & 17.86 & $-19.52$ & \nodata & \nodata \\
1.05 & 0.020 & 16.67 & $-19.37$ & 0.69 & 0.83 & 17.71 & $-19.52$ & 0.67 & 0.47 \\
1.05 & 0.025 & 16.05 & $-19.37$ & 1.69 & 0.59 & 17.34 & $-19.53$ & 1.59 & 0.21 \\
1.05 & 0.030 & 15.72 & $-19.38$ & 2.50 & 0.62 & 17.21 & $-19.54$ & 2.38 & 0.16 \\
1.05 & 0.035 & 15.58 & $-19.40$ & 2.85 & 0.64 & 16.98 & $-19.55$ & 2.71 & 0.17 \\
1.05 & 0.040 & 15.43 & $-19.41$ & 3.30 & 0.59 & 16.93 & $-19.57$ & 3.16 & 0.12 \\
1.05 & 0.045 & 15.31 & $-19.43$ & 3.63 & 0.56 & 16.72 & $-19.58$ & 3.49 & 0.10 \\
1.05 & 0.050 & 15.26 & $-19.44$ & 3.85 & 0.52 & 16.60 & $-19.58$ & 3.72 & 0.07 \\
1.05 & 0.060 & 15.21 & $-19.45$ & 4.00 & 0.38 & 16.57 & $-19.59$ & 3.74 & $-0.02$ \\
1.05 & 0.080 & 15.22 & $-19.48$ & 4.43 & 0.20 & 16.45 & $-19.61$ & 3.81 & $-0.14$ \\
\hline
1.10 & 0.010 & 16.82 & $-19.59$ & \nodata & \nodata & 17.51 & $-19.72$ & \nodata & \nodata \\
1.10 & 0.015 & 16.60 & $-19.58$ & 0.63 & 0.80 & 17.44 & $-19.71$ & 0.66 & 0.39 \\
1.10 & 0.020 & 16.08 & $-19.58$ & 1.82 & 0.43 & 17.13 & $-19.72$ & 1.69 & 0.06 \\
1.10 & 0.025 & 15.92 & $-19.59$ & 2.38 & 0.51 & 17.05 & $-19.72$ & 2.28 & 0.04 \\
1.10 & 0.030 & 15.79 & $-19.60$ & 2.77 & 0.51 & 16.89 & $-19.73$ & 2.67 & 0.05 \\
1.10 & 0.035 & 15.74 & $-19.61$ & 3.14 & 0.46 & 16.91 & $-19.74$ & 3.00 & 0.01 \\
1.10 & 0.040 & 15.69 & $-19.61$ & 3.36 & 0.41 & 16.86 & $-19.74$ & 3.19 & $-0.02$ \\
1.10 & 0.045 & 15.74 & $-19.62$ & 3.44 & 0.35 & 16.84 & $-19.75$ & 3.26 & $-0.05$ \\
1.10 & 0.050 & 15.76 & $-19.62$ & 3.55 & 0.28 & 16.80 & $-19.75$ & 3.26 & $-0.10$ \\
1.10 & 0.060 & 15.79 & $-19.63$ & 3.71 & 0.18 & 16.86 & $-19.75$ & 3.25 & $-0.16$ \\
1.10 & 0.080 & 15.77 & $-19.63$ & 4.24 & 0.07 & 16.88 & $-19.75$ & 3.47 & $-0.23$ \\
\enddata
\tablecomments{The synthetic observables are parametrized by the initial WD total mass and shell mass. The time since explosion for the $B$-band and $g$-band peak brightness is inversely related to the shell mass. The magnitude of the peak brightnesses is largely set by the total mass. The time since explosion of the early reddening maximum is characterized in Equation \ref{eq:time_model}. The magnitude of the reddening trends bluer as more radioactive decay heats the thicker shell models. Ellipses in the reddening columns denote models that are thin shell double detonations.}
\end{deluxetable*}

\bibliography{main}{}
\bibliographystyle{aasjournal}



\end{document}